\documentclass{article}
\usepackage{orcidlink}
\usepackage{amsmath,amssymb,amsfonts}
\usepackage[margin=4cm]{geometry}
\usepackage{algpseudocode}
\usepackage{algorithm}
\usepackage{algorithmicx}
\usepackage{arxiv}
\usepackage[utf8]{inputenc} 
\usepackage[T1]{fontenc}    
\usepackage{hyperref}       
\usepackage{url}            
\usepackage{graphicx}
\usepackage{plantuml}
\usepackage{multirow}
\usepackage{booktabs}       
\usepackage{flafter}		
\usepackage{longtable}
\usepackage{acronym}
\usepackage[acronym]{glossaries}
\usepackage{diagbox}

\graphicspath{ {./images/} }
\usepackage[acronym]{glossaries}
\title{ TableGuard - Securing Structured \& Unstructured Data }

\author{ 
    \href{https://orcid.org/0009-0004-1722-0306}
    {\includegraphics[scale=0.06]{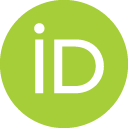}\hspace{1mm}Ajinkya Deshmukh$^*$}\\
	The A-Team (Pune, India)\\
    \texttt{Ajinkya.Deshmukh@synechron.com}\\
    \And
    \href{https://orcid.org/0000-0002-9064-3362}
    {\includegraphics[scale=0.06]{orcid}\hspace{1mm}Anantha Sharma$^*$}\\
	The A-Team (Charlotte, USA)\\
    \texttt{Anantha.Sharma@synechron.com}
}
     \renewcommand{\headeright}{}
 \renewcommand{\undertitle}{Synechron AI Practice}

\begin{document}

\maketitle

\def\thefootnote{*}\footnotetext{These authors contributed equally to this work.}

\begin{abstract}

With the increasing demand for data sharing across platforms and organizations, ensuring the privacy and security of sensitive information has become a critical challenge. 
This paper introduces “TableGuard”. An innovative approach to data obfuscation tailored for documents \& relational databases. 
Building on the principles and techniques developed in prior work \cite{2023arXiv230509550D} on context-sensitive obfuscation, 
TableGuard applies these methods to ensure that API calls return only obfuscated data, thereby safeguarding privacy when sharing data with third parties.

TableGuard leverages advanced context-sensitive obfuscation techniques to replace sensitive data elements with contextually appropriate alternatives. 
By maintaining the documents \& relational integrity and coherence of the data, our approach mitigates the risks of cognitive dissonance \cite{liu2023cognitivedissonancelanguagemodel} and data leakage. 
We demonstrate the implementation of TableGuard using a BERT based transformer model, which identifies and obfuscates sensitive entities within documents \& relational tables.

Our evaluation shows that TableGuard effectively balances privacy protection with data utility, minimizing information loss while ensuring that the obfuscated data remains functionally useful for downstream applications.  The results highlight the importance of domain-specific obfuscation strategies and the role of context length in preserving data integrity.

The implications of this research are significant for organizations that need to share data securely with external parties.  TableGuard offers a robust framework for implementing privacy-preserving data sharing mechanisms, thereby contributing to the broader field of data privacy and security.

\end{abstract}

\section{Introduction}

TableGuard aims to mask or replace sensitive information with fictitious but plausible data, thus preventing unauthorized access while retaining the utility of the dataset. However, indiscriminate obfuscation can lead to inconsistencies, especially when related entities are not obfuscated together. For example, replacing “New York” with “Chicago” while referring to the “Empire State Building” can confuse \cite{liu2023cognitivedissonancelanguagemodel} language models, leading to erroneous inferences. Our research focuses on obfuscating related entities cohesively to avoid such issues.

Traditional methods of data obfuscation often lead to significant information loss or cognitive dissonance  within automated systems, such as language models. This paper addresses these challenges by proposing a context-sensitive obfuscation approach that maintains document integrity. The paper is structured to detail our methodology, data preparation, model testing, and the implications of our findings, providing a comprehensive overview of effective PII obfuscation techniques.

\section{Data}

Our study uses names-dataset (a pypi package) \cite{nams_dataset_2023} which contains 730K first names, 983K last names - extracted from the Facebook massive dump (of 533M users). The composition of sensitive information like first name and last name coupled with dates of birth, and places of birth gives us a unique perspective on which names were popular around what year, we use this relationship to determine next best token to use in obfuscation.

This dataset provides a substantial foundation for testing the robustness and efficacy of our obfuscation techniques. Ensuring the ethical use and anonymization of data is paramount.

\section{Methodology}

The initial step in data preparation involved cleaning and preprocessing the dataset to remove any inconsistencies or irrelevant information. This process included normalizing the text, removing duplicates, and handling missing values. Each entry was then tokenized to facilitate further processing by our transformer model. Special attention was given to retaining contextual information to support accurate entity recognition and obfuscation.

Subsequently, named entity recognition (NER) was performed using Stanford CoreNLP \cite{conf/acl/2014-d}, a powerful NLP library, to identify entities that required obfuscation. These entities included names, locations, and other PII. The identified entities were then marked for replacement with appropriate pseudonyms or fictitious data. The preprocessing phase ensured that the dataset was ready for the application of our obfuscation model, setting the stage for effective and context-sensitive obfuscation.

\subsection{Approach}
Building on the work \cite{clark2016improvingcoreferenceresolutionlearning} we use a transformer-based model to identify sensitive entities within documents \& relational tables. The model is built on BERT \cite{devlin2019bertpretrainingdeepbidirectional} trained on a dataset of documents \& relational tables, where each table is represented as a sequence of tokens. The goal is to predict the replacement token for each entity in the table, given its context (context here is based on the content of the row in question coupled with data dictionary and table metadata).

\textbf{Masking}: 

Mask sensitive data by replacing characters or digits with placeholders. 
This experiment can assess the effectiveness of masking techniques in preventing unauthorized access to sensitive information.

\textbf{Example}:

    \begin{table}[ht]
        \centering
        \caption{Masking Example}
        \begin{tabular}{|c|c|c|}
            \hline
            \textbf{Original Text} & \textbf{Masked Text} & \textbf{Notes} \\
            \hline
            555.192.9277 & 555.XXX.XXXX & Phone number masking (using regex) \\
            \hline
            5423 3428 2372 9072 & 5XX3 XXXX XXXX 9072 & Credit card masking (using regex) \\
            \hline
            123 Any Street, Canada City, Canada & XXX Any Street, Canada City, Canada & Address masking (using NER \& regex) \\
            \hline
        \end{tabular}
        \label{tab:example_masking}
    \end{table}

These techniques are commonly used in data masking, but they do not take into account the context of the data.

\textbf{Perturbation}: Introduce noise or perturbations to sensitive data. There is a need to evaluate the impact of perturbation on data privacy and the ability to recover original data.

\textbf{Example}:

    \begin{table}[ht]
        \centering
        \caption{Perturbed Example}
        \begin{tabular}{|c|c|c|}
            \hline
            \textbf{Original Data} & \textbf{Perturbed Data} & \textbf{Notes} \\
            \hline
            12.34 & 12.39 & (gaussian noise) $\sigma = 0.1$ \\
            \hline
        \end{tabular}
        \label{tab:example_pertubation}
    \end{table}

\textbf{Differential Privacy}: Apply differential privacy mechanisms to the database.

\textbf{Example}:

\begin{table}[ht]
    \centering
    \caption{Differential Privacy Example}
    \begin{tabular}{|c|c|c|}
        \hline
        \textbf{Original Data} & \textbf{Differentially Private Data} & \textbf{Notes} \\
        \hline
        12.34 & 12.65 & (Laplace noise) $\epsilon = 0.5$ \\
        \hline
    \end{tabular}
    \label{tab:example_differential_privacy}
\end{table}

Various data engineering tools provide different data masking, differential privacy, and perturbation techniques.

\textbf{Contextual Obfuscation}: Obfuscate data based on contextual information. This experiment can explore the effectiveness of context-sensitive obfuscation in preserving data integrity and coherence.

\textbf{Example}:
This example describes FNOL (First Notice of Loss)

Orignal text:

\fboxrule=1pt
\fcolorbox{magenta}{lightgray}{
\resizebox{\dimexpr\linewidth-\fboxsep-\fboxrule}{!}{

    \parbox{\linewidth}{
On a rainy Tuesday morning,
Homer Simpson was driving to work when his car skidded on the wet pavement and collided with a lamppost near 789 Spooner Street, Springfield, IL 62629.
Shaken but unharmed, Homer immediately called his insurance company at (555) 555-1234 to report the accident. 
He provided his policy number, AB19010721, and explained the situation to the representative, Beth Sanchez. 

Beth, who was working from her office at 240 3rd St, Oakland, CA 94607, assured Homer that a tow truck would arrive shortly. 
She also sent a confirmation email to homer@mrplow.com, detailing the steps for filing the FNOL and scheduling an inspection. 
Homer, whose driver’s license number was WILR123456, felt relieved knowing that the process was underway and that he would soon receive assistance.    
    }
}
}

obfuscated (Context sensitive) text:

\fboxrule=1pt
\fcolorbox{blue}{white}{
\parbox{\linewidth}{
On a rainy \fcolorbox{green}{white}{Monday} \fcolorbox{red}{white}{Tuesday} morning,
\fcolorbox{green}{white}{Paul Buchman} \fcolorbox{red}{white}{Homer Simpson} was driving to work when his car skidded on the wet pavement and collided with a lamppost near 789 Spooner Street, Springfield, IL 62629.
Shaken but unharmed, \fcolorbox{green}{white}{Paul} \fcolorbox{red}{white}{Homer} immediately called his insurance company at \fcolorbox{green}{white}{(555) XXX-XXXX} \fcolorbox{red}{white}{(555) 555-1234} to report the accident. 
He provided his policy number, \fcolorbox{green}{white}{AB19XXXXX1} \fcolorbox{red}{white}{AB19010721}, and explained the situation to the representative, \fcolorbox{green}{white}{Annie Edison} \fcolorbox{red}{white}{Beth Sanchez}. 

\fcolorbox{green}{white}{Annie} \fcolorbox{red}{white}{Beth}, who was working from her office at 240 3rd St, Oakland, CA 94607, assured \fcolorbox{green}{white}{Paul} \fcolorbox{red}{white}{Homer} that a tow truck would arrive shortly. 
She also sent a confirmation email to \fcolorbox{green}{white}{xxxxx@xxxxxx.com} \fcolorbox{red}{white}{homer@mrplow.com}, detailing the steps for filing the FNOL and scheduling an inspection. 
Homer, whose driver’s license number was \fcolorbox{green}{white}{WXXXXXXX56} \fcolorbox{red}{white}{WILR123456}, felt relieved knowing that the process was underway and that he would soon receive assistance.
}
}

\begin{figure}[h]
    \centering
    \includegraphics[width=350pt]{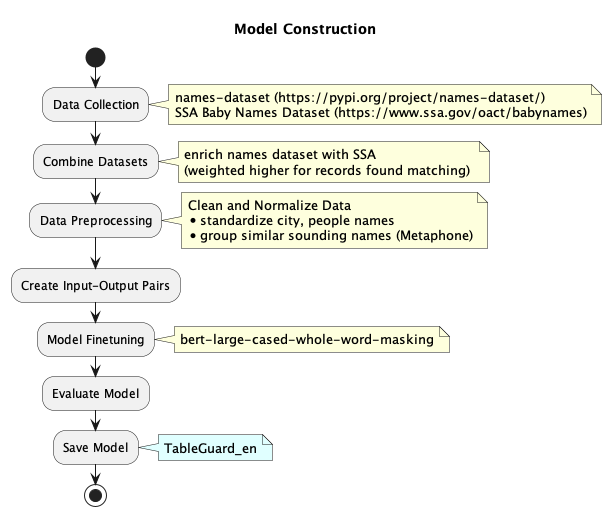}
    \caption{Model Construction}
    \label{fig:architecture}
\end{figure}

\section{Experiments}
Our experimental design aimed to rigorously evaluate the effectiveness of TableGuard's obfuscation techniques, focusing on their impact on data privacy and utility. The experiments were meticulously crafted to address key research questions regarding the trade-offs between privacy preservation and data usability while keeping a close eye on runtime performance.

\subsection{Setup}
The core of our experimental setup revolved around three primary obfuscation techniques: masking, perturbation, and differential privacy. Each technique was applied to a subset of the dataset, allowing us to isolate and measure their individual effects. The dataset, derived from the names-dataset package \cite{nams_dataset_2023}, provided a rich source of personally identifiable information (PII), including names, dates of birth, and places of birth, which are critical for our analysis.

The dataset used for the experiments was derived from a subset of the Facebook user dump, containing names, dates of birth, places of birth, and other PII. Before experimentation, the data was preprocessed to remove duplicates, handle missing values, and normalize text. 
Named entities were tagged using Stanford CoreNLP.

\subsection{Approach}
For each obfuscation technique, we implemented a series of tests to quantify the impact on both data privacy and utility. Privacy was assessed through measures such as information entropy and k-anonymity.

\subsection{Validation Methodology}
We tested the utility of the obfuscated data by running it through a series of common data analytics tasks, such as summarization, and trend analysis. We then compared the results with those obtained using the original (non-obfuscated) dataset.

\subsection{Metrics}
Task Performance was measured as F1 score of models trained on obfuscated versus original data.
Information Loss quantified as the percentage reduction in model performance after obfuscation.

\subsection{Results and Analysis}
The data utility experiments revealed that context-sensitive obfuscation resulted in minimal performance degradation (see graphs below). This highlights the effectiveness of context-sensitive techniques in preserving data utility while protecting privacy.

Our results revealed compelling insights into the effectiveness of each obfuscation technique. 
Masking, for instance, significantly reduced the readability of PII without compromising the overall structure of the data. 
Perturbation introduced noise that preserved the distribution of data points, albeit at the cost of increased uncertainty. 
Differential privacy offered a balance between privacy and utility, though at the expense of some information loss.

\begin{table}[ht]
    \centering
    \caption{Data load times}
    \begin{tabular}{|c|c|c|}
        \hline
        \textbf{Row Count} & \textbf{With Obfuscation Load time} & \textbf{No Obfuscation Load time} \\
        \hline
        100 & 50.169 & 0.118 \\
        \hline
        1,000 & 51.182 & 0.165 \\
        \hline
        10,000 & 53.161 & 2.513 \\
        \hline
        100,000 & 77.465 & 31.404 \\
        \hline
        150,000 & 91.724 & 49.141 \\
        \hline
        200,000 & 105.85 & 69.03 \\
        \hline
        300,000 & 132.436 & 106.357 \\
        \hline
        400,000 & 158.78 & 136.135 \\
        \hline
        500,000 & 183.628 & 167.247 \\
        \hline
        600,000 & 215.335 & 197.719 \\
        \hline
        700,000 & 248.337 & 233.751 \\
        \hline
        800,000 & 250.736 & 237.124 \\
        \hline
    \end{tabular}
    \label{tab:example_differential_privacy}
\end{table}

\begin{figure}[h]
    \centering
    \includegraphics[width=400pt]{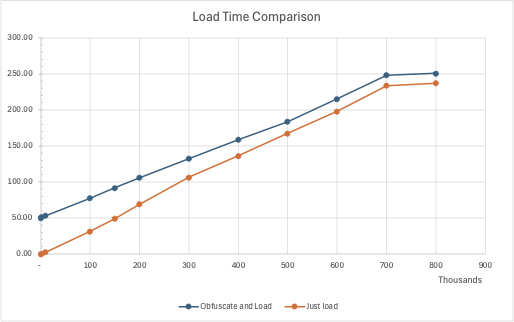}
    \caption{Load times}
    \label{fig:architecture}
\end{figure}

\pagebreak

\subsection{Implications}

The implications of our research extend beyond the immediate application of TableGuard. By demonstrating the feasibility of context-sensitive obfuscation, we contribute to the broader discourse on data privacy and security. 
Our work suggests that organizations can leverage sophisticated obfuscation techniques to share data securely, thereby fostering trust and collaboration among different stakeholders.

\section{Conclusion}
Our experimental evaluations demonstrated the effectiveness of TableGuard in balancing privacy protection with data utility, 
by minimizing information loss and ensuring that obfuscated data remains functionally useful for downstream applications, 
TableGuard showcases the importance of domain-specific obfuscation strategies and the critical role of context length in preserving data integrity. 
These findings underscore the potential of TableGuard as a robust framework for implementing privacy-preserving data sharing mechanisms, 
offering significant benefits for organizations seeking to share data securely with external parties.

The outcomes of our experiments underscore the complexity of achieving optimal data obfuscation. 
While masking and perturbation offer straightforward solutions, they come with trade-offs in terms of information loss and the potential for data leakage. 
Differential privacy, on the other hand, provides a more nuanced approach, allowing for controlled degradation of data quality in exchange for enhanced privacy guarantees.

Furthermore, our findings highlight the critical role of context in obfuscation. Context-sensitive obfuscation represents a promising direction for future research, offering the potential to mitigate the risks of cognitive dissonance and data leakage without sacrificing data utility.

\pagebreak

\section{Future Work}
Explore scalability and efficiency of TableGuard in real-world scenarios, including large-scale enterprise databases and cross-border data transfers (We have explored with nearly 1 million rows of data)

Exploring the legal and regulatory implications of using context-sensitive obfuscation techniques

Reducing complexity in complex legal documents

Securing medical reports and records (HIPPA compliance)

Securing MNPI

Added layer of security for documents in transit (in addition to the usual security measures)

Role \& Recipient based de-obfuscation of documents in popular communication channels (like email, Teams Chat etc.,)

\section{Acronyms}
NER = Named Entity Recognition

FNOL = First Notice of Loss

BERT = Bidrectional Encoder Respresentations from Transformers

HIPPA = Health Insurance Portability and Accountability Act

MNPI = Material Non Public Information

\bibdata{references}
\bibliographystyle{unsrt}
\bibliography{references}

\begin{thebibliography}{1}

\bibitem{2023arXiv230509550D}
Ajinkya {Deshmukh}, Saumya {Banthia}, and Anantha {Sharma}.
\newblock {Life of PII -- A PII Obfuscation Transformer}.
\newblock May 2023.

\bibitem{liu2023cognitivedissonancelanguagemodel}
Kevin Liu, Stephen Casper, Dylan Hadfield-Menell, and Jacob Andreas.
\newblock Cognitive dissonance: Why do language model outputs disagree with internal representations of truthfulness?
\newblock 2023.

\bibitem{nams_dataset_2023}
PyPI.
\newblock Names-dataset.
\newblock 2023.
\newblock Accessed: 2023-05-01.

\bibitem{conf/acl/2014-d}
Christopher~D. Manning, Mihai Surdeanu, John Bauer, Jenny~Rose Finkel, Steven Bethard, and David McClosky.
\newblock The stanford corenlp natural language processing toolkit.
\newblock In {\em ACL (System Demonstrations)\/} \cite{conf/acl/2014-d}, pages 55--60.

\bibitem{clark2016improvingcoreferenceresolutionlearning}
Kevin Clark and Christopher~D. Manning.
\newblock Improving coreference resolution by learning entity-level distributed representations, 2016.

\bibitem{devlin2019bertpretrainingdeepbidirectional}
Jacob Devlin, Ming-Wei Chang, Kenton Lee, and Kristina Toutanova.
\newblock Bert: Pre-training of deep bidirectional transformers for language understanding, 2019.

\end{thebibliography}

\end{document}